# Realization of higher coordinated Er in high-pressure cotunnite phase of $Er_2Ti_2O_7$

M. Modak[1,$], Rahul Kaiwart[1,2,$], Santosh K. Gupta[2,3], A. Dwivedi[1], K. K. Pandey[1,2], A. K. Poswal[4], and H. K. Poswal[1,2]*

[1]*High-pressure & Synchrotron Radiation Physics Division, Bhabha Atomic Research Centre, Mumbai 400085, India.*
[2] *Homi Bhabha National Institute, BARC Training School Complex, Anushaktinagar, Mumbai 400094, India.*
[3]*Radiochemistry Division, Bhabha Atomic Research Centre, Mumbai 400085, India.*
[4]*Atomic & Molecular Physics Division, Bhabha Atomic Research Centre, Mumbai-400085, India*

*Corresponding author: *himanshu@barc.gov.in*
$ *These authors contributed equally*

In this article we report the structural stability of $Er_2Ti_2O_7$ cubic pyrochlore with pressure using x-ray diffraction, Raman spectroscopy, photoluminescence, x-ray absorption and *ab-initio* calculations. Our studies establish a phase transformation in $Er_2Ti_2O_7$ from ambient cubic phase to high-pressure orthorhombic (cotunnite) phase, initiated at ~40 GPa. The transformation is sluggish and it does not complete even at the highest measured pressure in our study *i.e.* ~60.0 GPa. This is further supported by the first principle calculations which reveal that cotunnite phase is energetically more stable than the ambient phase above ~53 GPa. After complete release of pressure, the high-pressure cotunnite phase is retained while the fraction of untransformed pyrochlore phase becomes amorphous. Furthermore, the EXAFS data of the recovered sample at $L_3$ edge of $Er^{3+}$ ion show an increase in the coordination number of cations from eight at ambient to nine in the high-pressure phase. The mechanism of structural transformation is explained in terms of accumulation of cation antisite defects and subsequent disordering of cations and anions in their respective sublattice. The amorphization of the pyrochlore phase upon release is interpreted as the inability of accommodating the point defects at ambient conditions, which are formed in the pyrochlore lattice under compression.

# INTRODUCTION

Rare-earth based pyrochlore titanate ($R_2Ti_2O_7$, R: rare-earth) compounds have been extensively explored for their exceptional physical properties, extreme variety of functionality as well as the technological concern. They are equally important for the immobilization of actinide-rich nuclear wastes into ceramic form, which is stable for storage, transportation, and disposal[1, 2]. Furthermore, pyrochlore lattice is considered to be an ideal luminescence host material because of its large band gap, moderate phonon vibrations, ability to accommodate dopants at both *A* and *B* sites in addition to high chemical, mechanical, radiation, and thermal stability[3, 4]. It is well established that the pyrochlore titanate structure belongs to the *Fd-3m* space group with $R_2Ti_2O_6O'$ stoichiometry; whereas, a cubic unit cell consists of $TiO_6$ octahedra and $R_4O'$ tetrahedra. In general, the magnetic rare-earth ions are located at the vertices of the $R_4O'$ tetrahedra in the structure. The architecture of corner-sharing tetrahedra and crystal-field effects constrain the magnetic moments of the rare-earth ions to align along the axis, connecting the centers of the two neighboring tetrahedra[5]. Due to this exceptional geometry, the overall spin system does not minimize its spin-spin interactions in the ground state, leading to geometrical frustrations[6-8], which plays a significant role in the novel properties of these frustrated titanates. Furthermore, existing competing interactions such as crystal field anisotropy, quantum fluctuations, etc., can also results in various complex ground states at very low temperatures. Some of the examples are spin-ice states as observed in $Ho_2Ti_2O_7$[9, 10] and $Dy_2Ti_2O_7$[11, 12], spin-liquid state and magnetoelastic excitations in $Tb_2Ti_2O_7$[13-16], highly frustrated dipolar Heisenberg antiferromagnet in $Gd_2Ti_2O_7$[7, 17, 18], XY antiferromagnet with coexisting short-range and long-range ordering in $Er_2Ti_2O_7$[19, 20], etc. However, in some instances, the presence of impurity phases, structural defects including stuffing, antisite defects, etc. posses a crucial role to exhibit multiple ground states in a single titanate compound[21-23].

Probing a material subjected to extreme conditions is fundamental for the studies in materials science, nuclear engineering, and geosciences. Pressure, an important thermodynamic parameter, can strongly affect the structures and can change the delicate balance between the existing interactions and may direct us to perceive various physical properties of the $R_2Ti_2O_7$ pyrochlore titanate family. For example, $Tb_2Ti_2O_7$ exhibits spin-liquid behavior below 70 mK[16], whereas, with an external pressure of 8.6 GPa, it enters to a mixed phase of spin-liquid and spin-ice at 1.5 K[13, 24], due to the distortion of the pyrochlore lattice with pressure which help to reduce the frustration and develop magnetic correlation.

Most of the members in the $R_2Ti_2O_7$ pyrochlore titanate series mainly undergo either cubic to monoclinic or cubic to orthorhombic phase transformation under pressure. Among them, cubic to monoclinic transformation have been reported for $Yb_2Ti_2O_7$ at 28.6 GPa[8], $Ho_2Ti_2O_7$ at 37 GPa, $Y_2Ti_2O_7$ at 42 GPa, and $Tb_2Ti_2O_7$ at 39 GPa[25, 26]. Whereas, some other candidates reported/ predicted to exhibit pressure induced structural transitions from cubic pyrochlore to orthorhombic cotunnite phases; such as $Eu_2Ti_2O_7$ at 42 GPa, $Dy_2Ti_2O_7$ at 40 GPa[27], $Gd_2Ti_2O_7$ at 43.6 GPa[28], $Gd_{1.5}Ce_{0.5}Ti_2O_7$ at 42 GPa[29]. The signature of structural transformation in the $R_2Ti_2O_7$ using XRD from pyrochlore to monoclinic (M) or cotunnite (C) phase is very weak and difficult to uniquely assign the structure. In addition the structural transformation to M or C-phase is very sluggish and none of these compounds have accomplished complete transformations upto their highest pressure of the experiments. The available high-pressure Raman studies were also inconclusive since both the probable high-pressure phases are highly disordered and therefore don't show any sharp Raman modes. Hence, the conclusion drawn about the high-pressure phases in such titanate compounds have ambiguities and requires further investigations using additional techniques to probe local structures of the high-pressure phase.

$Er_2Ti_2O_7$ is an antiferromagnetic insulator with a negative Curie-Weiss temperature ($\theta_{CW}$ = -22 K) having ordering temperature (Neel Temperature: $T_N$ =) 1.2 K[19, 20]. Its ground state has been analyzed by considering it as an *XY* antiferromagnet on a frustrated lattice of corner-sharing regular tetrahedra with coexisting short- and long-range ordering[20]. However, in the ordering region, soft collective modes in the spin wave have been observed, which are associated with the unusual value of the canting angle between the antiferromagnetic spins[30-32]. Additionally, the magnetic ordering in the low temperature region is an example of the order-by-quantum disorder mechanism, in which the quantum fluctuations lift the degeneracy of the ground state, and tends to an ordered state[33-35]. A structural deformation due to a local rearrangement of the lattice without altering the cubic symmetry of the unit cell has also been reported in $Er_2Ti_2O_7$ below 130 K [36, 37]. However, the detailed structural investigation of $Er_2Ti_2O_7$ has not been explored under extreme condition of pressure.

It is therefore important to study the structural stability of $Er_2Ti_2O_7$ pyrochlore lattices with pressure to observe its ability to accommodate lattice disorder and to explore the possible high-pressure phase transitions. In general the structural transformations exhibited by the titanate pyrochlore compounds are too weak to assign the high-pressure phase with the help of only x-ray diffraction and Raman scattering experiment. We have therefore, performed systematic investigation of structural response of polycrystalline form of $Er_2Ti_2O_7$ pyrochlore using x-ray diffraction, Raman scattering, photoluminescence and x-ray absorption experiments. These combined studies with insights for both local and long-range structural details help establish the high-pressure phase of $Er_2Ti_2O_7$ more reliably. The observed high-pressure behavior, validated with our *Ab-initio* calculations is explained in terms of defect formation energy.

## EXPERIMENTAL DETAILS

Polycrystalline $Er_2Ti_2O_7$ samples were synthesized by solid-state reaction, starting with high purity powders of Erbium oxide ($Er_2O_3$, 99.9% Sigma-Aldrich) and Titanium dioxide ($TiO_2$, 99.0% Hi-media) in stoichiometric ratios through ball milling and respective intermediate heating. Further, the details of sample preparation is provided elsewhere[38]. High-pressure (HP) x-ray diffraction (XRD) experiments have been performed employing diamond anvil cell (DAC) at the Extreme Conditions x-ray diffraction (ECXRD) beamline (BL-11) at Indus-2, RRCAT, India using monochromatic x-rays of wavelength, $\lambda$ = 0.5466 Å. Pressure in sample was estimated using the equation of state of gold, loaded in the DAC along with sample and pressure transmitting medium.

Raman spectra and photoluminescence (PL) with high-pressure were collected in Jobin Yvon triple-stage T64000 Raman spectrometer in a back scattering geometry and single stage mode of operation. Argon ion Laser of wavelength 488 nm has been used as excitation source. For high-pressure generation, symmetric type DAC coupled with the gas membrane has been used for all the experiments. During HP Raman and PL experiments, pressure has been calibrated using the Ruby fluorescence method[39] and first order Raman mode of diamond at the centre of the culet[40, 41]. The sample along with few particles of gold and Ruby chip were loaded (as pressure marker) in the sample chamber of ~ 100 μm , created on a pre-indented tungsten gaskets for HP XRD and HP Raman/PL experiments, respectively. For all the HP experiments, hydrostatic conditions (upto 10.5 GPa) inside the sample chamber were maintained using a mixture of methanol and ethanol (4:1)[42], acting as the pressure transmitting medium (PTM).

X-ray absorption spectroscopy (XAS) measurements have also been carried out at BL-11, Indus-2 in continuous-energy-scan mode. The details of beamline, measurement protocols and related parameters are as described in our earlier report[43]. X-ray absorption near edge structure (XANES) and extended x-ray absorption fine structure (EXAFS) have been recorded at the $L$ edges ($L_2 \rightarrow L_3$) of Er atom. All data were processed using Athena software, and EXAFS data analysis was performed using FEFF6 and the Artemis program[44, 45].

## COMPUTATIONAL DETAILS

Density functional theory (DFT) based first principles calculations have been carried out to estimate the total energy of the system as implemented in Quantum Espresso package 6.0. Projected augmented wave (PAW) technique is used for self-consistency field (SCF) calculation. For the exchange and correlation energy, generalize gradient approximation (GGA) scheme has been considered. Pseudo-potentials for Erbium, Titanium, and oxygen atoms have been taken from PSlibrary. The kinetic energy and charge density cut-off in the plane wave basis set expansion is taken to be 80 Ry and 800 Ry respectively. The energy convergence criterion is set to $< 1 \times 10^{-8}$ Ry. The Monkhorst–Pack grid of $3 \times 3 \times 3$ k-points has been taken for k-point integration in the Brillouin zone. The geometric structures have been optimized using Broyden–Fletcher–Goldfarb–Shanno (BFGS) minimization scheme by minimizing the forces on individual atoms below $10^{-2}$ eV/Å. All the calculations have been performed by treating 4f electrons of Er atom as the core electrons as they weakly influence the phase transition pressure, particularly in rare-earth titanates[28]. Simulations have been performed by considering 11 valence electrons of the Er atom ($5s^2 6s^{1.5} 5p^6 6p^{0.5} 5d^1$), 12 electrons of the Ti atom ($3s^2 4s^2 3p^6 3d^2$), and 6 electrons of the O atom ($2s^2\ 2p^4$). Structure has been optimized for $Er_2Ti_2O_7$ in pyrochlore structure (space group *Fd-3m*) and cotunnite phase (space group *Pnma*) under ambient condition and with pressure upto 56 GPa in the step of 2 GPa

## RESULTS AND DISCUSSIONS

### HIGH-PRESSURE X-RAY DIFFRACTION

The XRD patterns of $Er_2Ti_2O_7$, collected at ambient condition are presented in Fig. 1(a). The pattern can be indexed with cubic pyrochlore phase (space group *Fd-3m*). Other strong peaks are identified as Gold (Au: pressure marker) and Tungsten (W: gasket material) and indicated accordingly. The pattern has been refined by Rietveld refinement method as implemented in GSAS-II software and the lattice parameter of the cubic pyrochlore structure is obtained as $a$ = 10.0916(4) Å.

The material is subjected to high-pressure and *in situ* XRD patterns have been recorded at intermediate steps while increasing pressure. The evolution of XRD patterns at few selective pressures are plotted in Fig. 1(b). Diffraction peaks due to W gasket and Au have been marked as * and #, respectively. It is ought to be mentioned that with increase of pressure, all the peaks shift towards higher Braggs angle, due to the compression of unit cell of the compound, as expected. However, no changes in the diffraction patterns except the right shift have been noticed upto ~40 GPa, indicating that the ambient cubic phase remains stable till then.

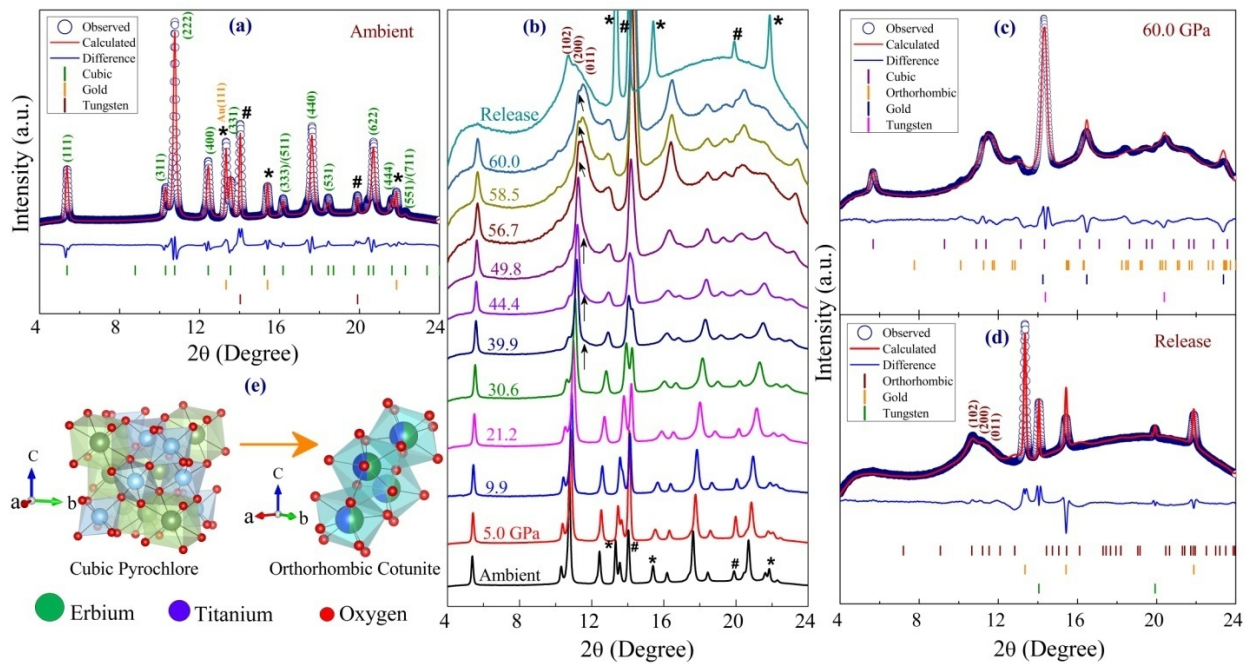

**Figure 1.** (a) Room temperature ambient XRD pattern with Rietveld refinement. Peaks due to gold and tungsten have been marked with * and #, respectively. (b) Ambient and high-pressure XRD patterns upto 60.0 GPa, black arrows (↑) indicating the formation of new phase above ~ 40 GPa. (c) Fitted XRD spectra at the highest pressure ~ 60.0 GPa considering the high-pressure orthorhombic cotunnite phase with the remaining ambient cubic phase. (d) Release Spectra, refined with irreversible high-pressure orthorhombic phase. (e) Refined unit cell of the ambient cubic pyrochlore structure and high-pressure orthorhombic structure.

With further increase of pressure, it has been observed that a new diffraction peak is evolved at ~11.5º, at the right side of the most intense ambient (222) peak, as indicated by ↑. Moreover, an additional peak at 11.3º is developing and getting prominent at higher pressure, above ~ 56 GPa (marked ↑). This suggests the structural rearrangement and movements of the atoms in the system leading to a structural transformation under pressure. The newly developed peaks have been recognized as a convolution of (102), (200) and (011) peaks of high-pressure orthorhombic cotunnite phase (space group *Pnma*). This result is consistent with many titanate pyrochlore compounds reported to undergo pyrochlore to cotunnite like transformation such as $Eu_2Ti_2O_7$ at 40 GPa, $Dy_2Ti_2O_7$ at 42 GPa[27], $Gd_2Ti_2O_7$ at 43.6 GPa[28], $Gd_{1.5}Ce_{0.5}Ti_2O_7$ at 42 GPa[29] and so on. The transformation does not complete even at the highest pressure of the experiment, i.e. ~ 60.0 GPa, and some fraction of ambient pyrochlore phase is still visible along with the high-pressure phase. The Rietveld refinement of the XRD pattern at the highest pressure of the experiment has been represented in Fig. 1(c) and the Bragg's angle of prominent peaks of (102), (200) and (011) reflections of the new phase are given as 2θ = 11.20º, 11.80º and 11.84º, respectively. Fractional coordinates have been taken from DFT calculations. The other reflections from the HP cotunnite phase might be weak and merged with the broadened peaks of the ambient phase/Au/W. The refined lattice parameters of the high-pressure orthorhombic phase at ~ 60.0 GPa are $a$ = 5.2693(1) Å, $b$ = 3.1948(9) Å, and $c$ = 6.3672(7) Å. After releasing pressure the transformed HP orthorhombic phase sustained, while the remaining untransformed ambient phase turned into an amorphous phase. The release spectra has been refined (as shown in Fig. 1(d)) and the obtained lattice parameter of the retained orthorhombic phase at ambient pressure are $a$ = 5.5990(1) Å, $b$ = 2.9574(2) Å, $c$ = 6.9135(7) Å. Unit cell structures of the ambient pyrochlore phase and the high-pressure cotunnite phase have been presented in Fig. 1(e).

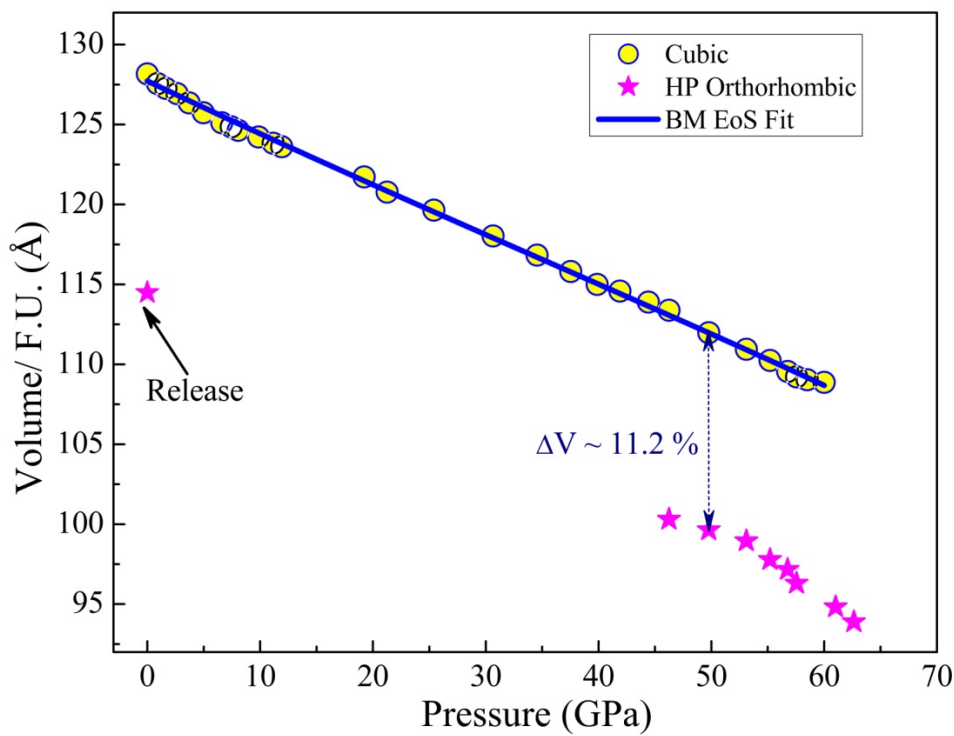

**Figure 2.** Variation of cell volume per formula unit of ambient cubic and HP orthorhombic phase obtained from the experiments.

At 50 GPa, the drop in volume of the high-pressure phase with respect to the ambient cubic phase has been experimentally attained as 11.2 %. The PTM used in our experiment *i.e.* methanol:ethanol mixture freezes above 10.5 GPa. Therefore, the bulk modulus has been determined within the hydrostatic limit of the PTM, by fitting the experimentally obtained *P-V* values to the third order Birch-Murnaghan equation of state (BM EoS) using EOSFit7 software[46] and the plot is presented in Fig. 2. From the fit, the bulk modulus ($B_0$) and the first order derivative of the bulk modulus ($B_0'$) are obtained as 220 (13) GPa and 4 (2) respectively. These values are comparable to the other titanate materials reported elsewhere.

Many authors have previously described the 'cotunnite-like' structure, as an overall increase in the coordination number of cations from six and eight in pyrochlore to eight, nine, or ten in the cotunnite-like structure[47, 48]. However, in the earlier studies the exact coordination number of cations has not been reported. The transformation from ordered pyrochlore to disordered cotunnite like structure requires cation and anion sublattice disordering in the system. Due to presence of disorder in high-pressure phase as well as large non hydrostatic stress under pressure, the diffraction data quality is very limited and additional spectroscopic analysis are necessary to further complement the observed phase transition. The mechanism of structural transformation has been discussed later using density functional theory calculation.

## HIGH PRESSURE RAMAN SCATTERING

Raman spectroscopy measurements have been performed to complement x-ray diffraction results which offer additional insights into the short-range ordering in the system. Group theoretical analysis for the pyrochlore titanates, having stoichiometry of $R_2Ti_2O_6O'$ (space group *Fd-3m*), predicts following vibrational normal modes, such as, six Raman active modes, as $A_{1g} + E_g + 4F_{2g}$; seven infrared active modes, as $7F_{1u}$; one acoustic mode and inactive modes: $2F_{1g} + 3A_{2u} + 3E_u + 4F_{2u}$[49-51]. The ambient temperature Raman spectra of $Er_2Ti_2O_7$ at ambient condition, collected using 488 nm laser excitation source, is presented in Fig. 3. The spectra can be fitted with 9 Lorentzian peaks, using least square method in Origin software, abbreviated with M1 to M9. In accordance with previous studies[36, 51, 52], M1 is recognized as PL band, M2 can be attributed to disorder induced Raman active mode and is generally associated with the interaction of Ti atom with *48f*-oxygen atoms. M3, M5 and M7 are identified as three $F_{2g}$ Raman modes and M4 and M6 are assigned as $E_g$ and $A_{1g}$ modes respectively. M6 and M7 vibrations are mainly related to Er-O bending modes. While M3 and M4 are primarily associated with with O–Er–O bending, Er–O stretching, and Ti–O bending vibrations. M8 and M9 are determined as two phonon process, which are also observed in other pyrochlore titanates. It has been evidenced that M5 is a combination of $F_{2g}$ phonon and PL line. The fourth $F_{2g}$ mode, which should appear near 700 cm$^{-1}$ in $Er_2Ti_2O_7$ is either missing, or could not be identified due to the broad M9 band. The disorder-induced Raman active mode, M2 is associated with the vibration of $Ti^{4+}$ atoms form a tetrahedral network with vacant 8a-sites at the center of each tetrahedron.

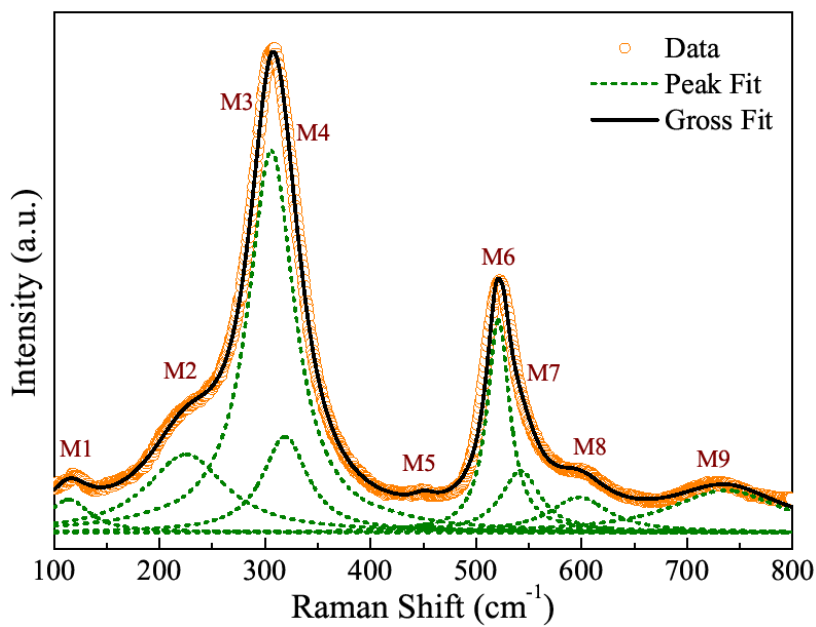

**Figure 3.** Fitted Raman spectra of $Er_2Ti_2O_7$ at ambient conditions with identified modes.

The evolutions of Raman signals at a few selective high-pressures have been represented in Fig. 4. With increase of pressure, the entire peaks shift towards higher energy, which is expected due to the lattice contraction. When pressure is increased to about ~37 GPa, the most prominent peaks due to ambient pyrochlore phase, namely M3 and M6 slowly weakens and their FWHM increase. At this pressure, a new peak is evolved adjacent to peak M3/M4 of the ambient pyrochlore phase which has been marked as '↑' in Fig 4. This is also apparent in Fig 5, where we have shown linear variation of all the observed modes. Mode M4 shows change in slope at ~37 GPa which is due to emergence of new peak corresponding to high-pressure phase. Meanwhile, the intensity of mode M8 is seen to slightly increase with compression, (marked ↑) in Fig. 4. When pressure is further increased to about ~45 GPa, the new peak gets broadened and disappears. With further compression to about ~51 GPa, all the modes corresponding to pyrochlore phase has been lost and only a single mode at about ~800 cm$^{-1}$ remains which exist upto the highest pressure of the experiment *i.e.* 58.6 GPa. Therefore, Raman scattering measurements agrees with the HP XRD measurement and suggest that ETO undergoes structural transformation at about ~ 37 GPa. The absence of any sharp peaks due to high-pressure phase indicates that it is a disordered structure. Indeed, the signature of Raman mode seems quite similar to the nature of HP cotunnite like phase, reported for other titanate compounds[27, 29].

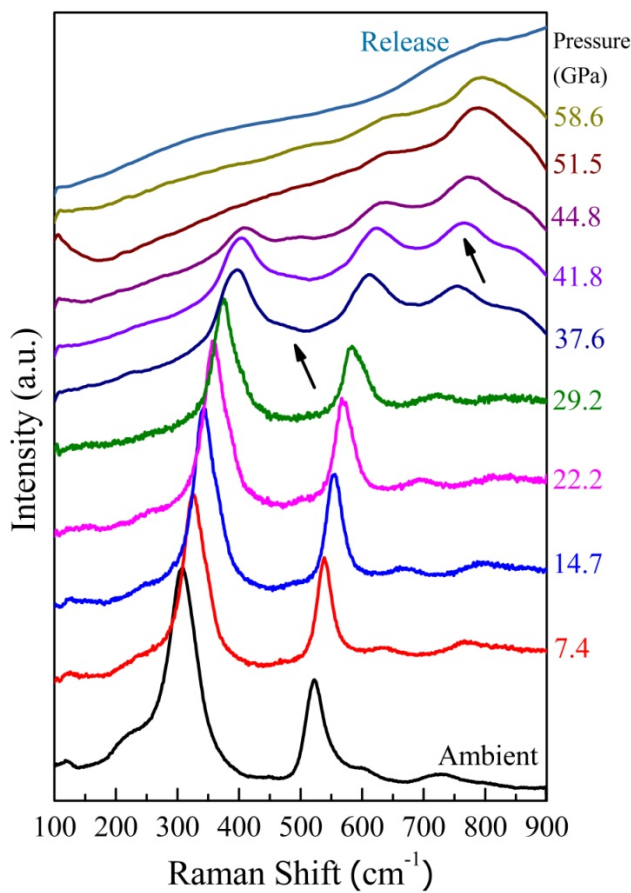

**Figure 4.** Pressure evolution of the Raman spectra and after releasing the pressure.

The intensity of Raman modes of the ambient phase has been completely vanished above 50 GPa, but, interestingly, at this pressure high-pressure x-ray diffractions still shows sharp Bragg peaks corresponding to the ambient pyrochlore phase in addition to the high-pressure phase. This suggests that though the ambient pyrochlore phase maintains long range ordering through x-ray diffraction, locally it is disordered mainly the positional disordering due to generation of point defects such cation antisite and anion Frenkel defects. This is probably the reason for absence of any sharp Raman modes due to the high-pressure phase, similar to the defect fluorite structure. The local disordering of pyrochlore lattice is starting point for structural transformation into high-pressure phase which itself has random distribution of cation and anion sublattice. Therefore high-pressure phase also doesn't support sharp Raman mode and only single prominent Raman mode appearing between 700−800 cm$^{-1}$ is observed which is believed to be due to distortion of the TiO$_6$ octahedra[50, 53]. The HP phase has been identified as orthorhombic cotunnite like structure from HP XRD studies.

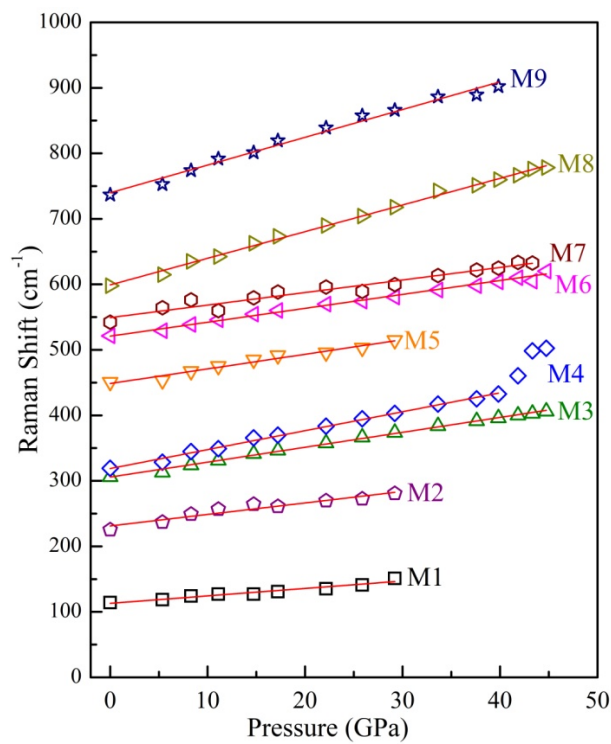

**Figure 5.** Shift of various peak positions with increasing pressure. Error bars in the peak positions, obtained in fitting, are inside the symbol. Red lines have been drawn using linear fit of the corresponding peak positions as a function of pressure.

## HIGH PRESSURE PHOTOLUMINISCENCE

Er$_2$Ti$_2$O$_7$ exhibits photoluminescence behavior due to presence of Er$^{3+}$ ions in its lattice. Emission spectrum at ambient and with pressure are plotted in Fig. 6 and Fig. 7, respectively. The ground state electronic configuration of Er$^{3+}$ ion is given by $^4I_{15/2}$. When Er$_2$Ti$_2$O$_7$ is excited by the argon ion laser of wavelength 488 nm (2.54 eV), electrons from the valence band of the host gets excited and populated in $^4F_{7/2}$ and electrons are subsequently non-radiatively de-excited into $^2H_{11/2}$, $^4S_{3/2}$ and $^4F_{9/2}$ level, each of which is split into $(J + \frac{1}{2})$ Kramers doublet levels due to stark effect. The radiative transition observed under ambient are then resulted from various transition taking place from $^2H_{11/2}$, $^4S_{3/2}$ and $^4F_{9/2}$ to the ground state of $^4I_{15/2}$, which splits into eight sub-levels. This splitting may give multiple lines as observed under ambient conditions. Among the observed PL bands, one is in the red region due to $^4F_{9/2} \rightarrow {}^4I_{15/2}$ transition and others are in the green region from $^2H_{11/2} \rightarrow {}^4I_{15/2}$ and $^4S_{3/2} \rightarrow {}^4I_{15/2}$ transitions. In this study we have only presented pressure variation of $^4S_{3/2} \rightarrow {}^4I_{15/2}$ band lying in the green region.

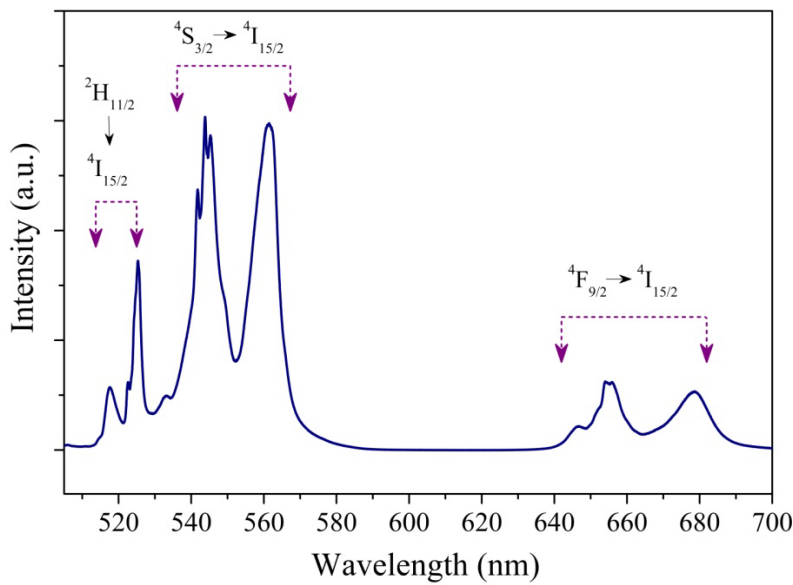

**Figure 6.** Ambient photoluminescence spectra of ETO indicating Green and Red transition bands.

When pressure is increased, the entire spectra slowly shift towards higher wavelength. The FWHM of peaks are also seen to broaden with pressure which is understood in terms of disordering in the structure under pressure inherent to the pyrochlore lattice of the host. Meanwhile, their respective intensities gradually decrease with pressure and at ~37.6 GPa, these bands have lost most of their intensities. At this pressure, a few new peaks emerge and their respective intensity also increases monotonously. The loss of the emission bands at ambient can be interpreted in terms of local site symmetry modification around $Er^{3+}$ ions in the host compound. Indeed from HP XRD, the compound undergoes pyrochlore to cotuunite like structural transformation above 40 GPa. During this change, the local site around $Er^{3+}$ changes its symmetry from $D_{3d}$ to $C_s$ leading to breakdown of many of the selection rules giving rise to more number of emission lines.

The new bands observed in the emission spectra is present due to the favorable pathways for radiative de-excitation after the electrons have been transferred into $^4F_{7/2}$ creating so called charge transfer band (CTB). The CTB is the property of the host material. Since the host material has undergone structural phase transformation, CTB modifies. The lack of the CTB then leads to the loss of the ambient emission bands. The structural transformation leads to the modification of the electronic structure of the materials leading to new CTB and therefore, the emission path ways change and consequently new emission lines appear. Further increase in pressure leads to shifting in the emission lines towards higher wavelength and this trend is observed until ~58.6 GPa, which is the highest pressure of the experiment. When the pressure is completely released from the sample, the observed pattern contain peaks corresponding to the high-pressure phase along with an envelope of a few broad peaks profile. The envelope probably corresponds to the amorphous phase due to remaining portion of ambient pyrochlore phase. The structural phase transformation is irreversible upon release of the pressure. High-pressure PL measurements thus support the finding of the x-ray diffraction and Raman spectroscopic measurement that system undergoes structural phase transformation at about ~37 GPa.

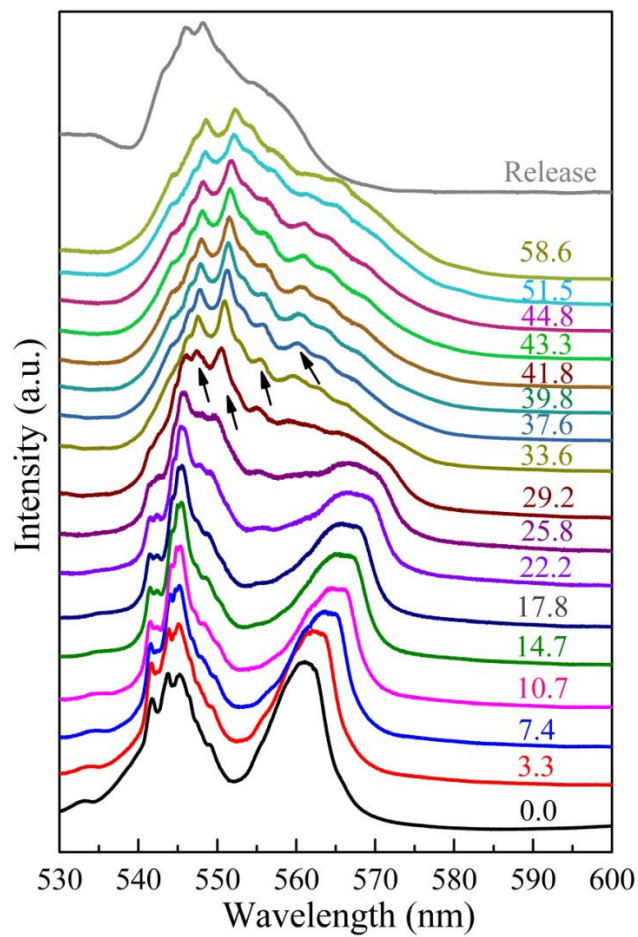

**Figure 7.** Evolution of PL spectra with increasing pressure and after releasing pressure.

## X-RAY ABSORPTION

In our discussion on high-pressure x-ray diffraction experiments, we have suggested that $Er_2Ti_2O_7$ undergoes pyrochlore to cotunnite phase transformation initiating at ~40 GPa. After pressure cycling to ~60.0 GPa, the high-pressure phase is retained upon complete release and from the Rietveld refinement we have interpreted that all the remaining sharp peaks may be indexed to the cotunnite phase (space group *Pnma*). In order to further complement our findings we have performed x-ray absorption measurements on the recovered sample after compression and compared it with the ambient pyrochlore phase.

The EXAFS signal $\chi(k)$ was obtained by pre-edge and post edge bare atomic background data subtraction from the measured absorption coefficient and normalizing the EXAFS oscillation by the edge step. The plot of $k^2\chi(k)$ is shown in Fig. 8(a). The pseudo radial distribution function around the absorbing embedded atom was obtained by Fourier transformation of the extracted signal using a Hanning window in the $k$ range of $2-11$ Å$^{-1}$. The moduli of the Fourier transform are also shown with pressure in Fig. 8(c).

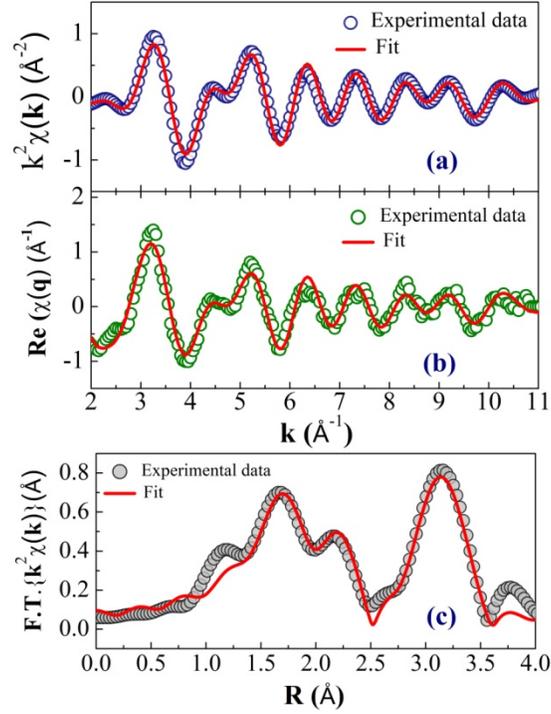

**Figure 8.** Comparison between the experimental Er $L_3$-edge EXAFS spectrum (shown by circles) at ambient and model calculation (shown by the solid curve) corresponding to a first and second coordination shell is reported for (a) the back-transformed signal, (b) the extracted $k^2\chi(k)$ signal, and (c) the moduli of the Fourier transforms.

A quantitative analysis of the EXAFS data at the $L_3$-edge of Er$^{3+}$ ion has been performed using the ARTEMIS package[44,45]. The calculation of backscattering amplitudes and phases was carried out using the FEFF6 code. For the local structure around Er atom in Er$_2$Ti$_2$O$_7$ at ambient, we have considered the known crystalline phase of Er$_2$Ti$_2$O$_7$ in the pyrochlore structural form (eight-coordinated). The local structure around Er atoms in the pyrochlore phase has a $(2+6)$−type double distance distribution of eight oxygen atoms in [Er−O] polyhedra. The EXAFS signal $\chi(k)$ shows strong contribution from the second shell coordination around Er atom as well. Therefore, we have taken both, the first and the second shell Er−O path in the modeling. For modeling of high-pressure phase of Er$_2$Ti$_2$O$_7$, for which an increase in the coordination number of Er is expected, has been tested by taking the Er − O bond distance with respect to the eightfold-coordinated pyrochlore model and setting different fixed coordination numbers. EXAFS data fittings have strongly coupled fitting parameters therefore all the modeling has been performed with fix number of coordination number. For the high-pressure orthorhombic phase around the Er$^{+3}$ ions, we have fitted the experimental data with four different models based on eightfold−, ninefold−, tenfold−, and elevenfold−coordination geometry. We have used reduced $\chi^2(\chi_v^2)$ as a relevant parameter to determine which of the models gave the best fit to the observed profile of pressure released Er$_2$Ti$_2$O$_7$. The amplitude reduction factor parameter $S_0^2$ (1.0) has been kept constant over all the fittings and taken from the ambient data fitting[54]. We have only refined the values of the Er − O bond distance and mean-square relative displacement $\sigma^2$. Fig. 8 represents the best fit between experimental data and the fitted curve using the eightfold-coordination model at ambient. Table I gives all the refined parameters for the ambient pyrochlore structure.

Table I. Refined parameters for the ambient pyrochlore structure of $Er_2Ti_2O_7$

| First Coordination Shell (Er−O) | | | |
|---|---|---|---|
| $N =$ | | 8 | |
| $N_1$ | $N_2$ | 6 | 2 |
| $\sigma^2$ | | 0.011 | 0.0002 |
| Er − O | | 2.38 ± 0.04 | 2.20 ± 0.04 |
| Second Coordination Shell (Er−Er) and (Er−Ti) | | | |
| | | Er − Er | Er − Ti |
| $N =$ | | 6 | 6 |
| Bond distance | | 3.57± 0.05 | 3.52± 0.02 |
| $\sigma^2$ | | 0.015 | 0.011 |
| $\chi_v^2$ | | 188 | |

Table II. Refined parameters for the high-pressure orthorhombic cotunnite structure of $Er_2Ti_2O_7$

| $N =$ | | 8 | | 9 | | 10 | | 11 | |
|---|---|---|---|---|---|---|---|---|---|
| $N_1$ | $N_2$ | 6 | 2 | 6 | 3 | 6 | 4 | 6 | 5 |
| $\sigma^2$ | | 0.013 | 0.005 | 0.017 | 0.007 | 0.022 | 0.009 | 0.027 | 0.01 |
| $Er - O$ | | 2.41±0.04 | 2.27±0.04 | 2.45±0.03 | 2.29±0.03 | 2.47±0.03 | 2.30±0.03 | 2.50±0.04 | 2.32±0.04 |
| $\chi_v^2$ | | 336 | | 203 | | 198 | | 246 | |

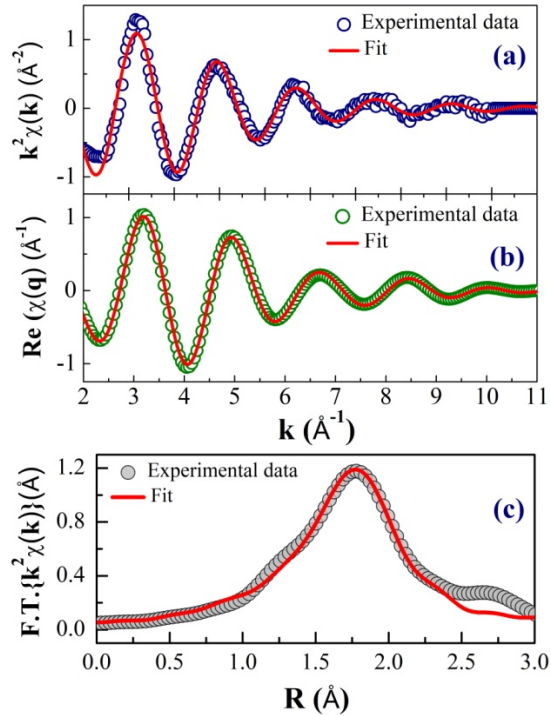

**Figure 9.** Comparison between the experimental Er $L_3$-edge EXAFS spectrum (shown by circles) at ambient after pressure release (*i.e.* recovered sample) and model calculation (shown by the solid curve) corresponding to a single coordination shell is reported for (a) the back-transformed signal, (b) the extracted $k^2\chi(k)$ signal, and (c) the moduli of the Fourier transforms.

Fig. 9 represents fitting corresponding to the recovered sample after decompression and Table II shows all the fitting parameters. It turns out that the ninefold− and tenfold−coordination models both give comparable best fitting to the experimental data points. However, for a comparison between experimental data and the best fit, the ninefold-coordination model has been shown in Fig. 9, because ninefold−coordination is generally accepted as the coordination number of cations in the high-pressure cotunnite phase[4].

In the next step, we want to discuss the evolution of the Er − O bond distance after pressure cycling. It is worth mentioning here that the Er − O bond length is not affected by the variation in the EXAFS oscillation amplitude, and hence it is unaffected by changes in the sample thickness under pressure[55]. From Table I, in the ambient pyrochlore phase of $Er_2Ti_2O_7$ six of the oxygen atoms are located at ∼2.38 Å, while remaining two are located at ∼2.20 Å. The average Er − O bond distance in $ErO_8$ polyhedra is then ∼2.33 Å. Second shell coordination refinement gives Er − Er at about ∼3.57 Å and Er − Ti at about ∼3.51 Å. For the high-pressure cotunnite phase, six oxygen atoms are located at ∼2.45 Å and remaining three oxygen atoms located at ∼2.29 Å. The average Er − O bond distance is about ∼2.40 Å.

To the best of our knowledge, this is the first x-ray absorption measurements on the high-pressure orthorhombic phase of any of the rare-earth titanates type of compounds. From these experiments we argue that due to structural transformation from pyrochlore phase to the orthorhombic phase the coordination number around Er increases from eight to nine while the Er − O bond distance increases from 2.33 Å to 2.40 Å at ambient conditions. This result is consistent with the *Ab initio* calculation on the other pyrochlore system undergoing transformation into cotunnite phase[56]. This jump is expected due to the volume discontinuity, which is a signature of the first-order structural transformation. Debye-Waller factor $\sigma^2$ values have also been shown in Table I and II. An increase in $\sigma^2$ is evident for the high-pressure orthorhombic phase, which is understood in terms of an increase in static disorder within the system arising due to the structural phase transformation to higher coordinated phase and higher Er − O bond distance.

## COMPUTATIONAL STUDIES

For understanding the high-pressure behavior of $Er_2Ti_2O_7$ and its phase stability, density functional theory (DFT) based first principle calculations have also been performed. We have calculated enthalpies of the two phases (1) Pyrochlore (2) Cotunnite structure with pressure from 0 to 56 GPa in the steps of 2 GPa. The cotunnite phase is a highly disordered phase where both the cations and anions are randomly distributed on *4c* positions and one eighth of the oxygen atoms are missing. All the simulations have been performed with a cell consisting of 88 atoms. For the present calculation we have used randomization method in order to create the defect cotunnite lattice. We have considered more than 40 different configurations consisting of random distribution of cations and anions on their sublattice which have been optimized and the most stable among these have been discussed here. In contrast with the defect cotunnite structure, pyrochlore is an ordered structure where Er and Ti occupy *16d* and *16c* positions respectively. Pyrochlore phase contains eight formula units in its unit cell while cotunnite phase contains one formula unit per unit cell. The optimized lattice parameters for pyrochlore phase at ambient are $a = 10.0826$ Å, $V = 1026.033$ Å$^3$ and for cotunnite phase at ambient are $a = 5.6229$ Å, $b = 2.9967$ Å, $c = 6.9880$ Å and $V = 117.423$ Å$^3$. The optimized fractional coordinates for pyrochlore and defect cotuunite phase have been provided in Table III and IV respectively.

Table III. Optimized fractional coordinates for ambient pyrochlore phase of $Er_2Ti_2O_7$.

| Pyrochlore phase (*Fd3m* #227) | | | |
|---|---|---|---|
| **Atom** | **Wyckoff position** | **Point Symmetry** | **Coordinates** |
| Er | *16d* | -3m ($D_{3d}$) | 0.5, 0.5, 0.5 |
| Ti | *16c* | -3m ($D_{3d}$) | 0, 0, 0 |
| O1 | *48f* | mm ($C_{2v}$) | 0.3424, 0.1250, 0.1250 |
| O2 | *8b* | -43m ($T_d$) | 0.3750, 0.3750, 0.3750 |

Table IV. Optimized fractional coordinates for high-pressure orthorhombic cotunnite phase of $Er_2Ti_2O_7$ at ambient.

| Cotunnite phase (*Pnma* #62) | | | |
|---|---|---|---|
| **Atom** | **Wyckoff position** | **Point Symmetry** | **Coordinates** |
| Er/Ti | *4c* | .m. ($C_s$) | 0.2520, 0.2500, 0.3839 |
| O | *4c* | .m. ($C_s$) | 0.0743, 0.2500, 0.6883 |
| O | *4c* | .m. ($C_s$) | 0.1346, 0.2500, 0.0647 |

Relative enthalpy (Fig. 10(a)) of these phases with respect to the ambient cubic $Er_2Ti_2O_7$ shows that the orthorhombic $Er_2Ti_2O_7$ gets stabilized at nearly 53 GPa. This agrees well with many other reports on various titatanates such as $Y_2Ti_2O_7$, $Dy_2Ti_2O_7$, $Gd_2Ti_2O_7$ and so on in a pressure range of 40–50 GPa[27, 28, 56, 57]. The value of phase transition pressure from pyrochlore to orthorhombic structure differs by +13 GPa compared to the experimental values. This may be due to not taking into account the effect of non-hydrostatic stress in the calculations. As the used PTM methanol:ethanol mixture solidifies above 10.5 GPa, the components of shear stress probably contribute strongly to bring down the phase transition pressure to ~40 GPa. During the phase transition there is a volume drop of ~9.1 %, which is in agreement with the experimental finding (~11.2%).

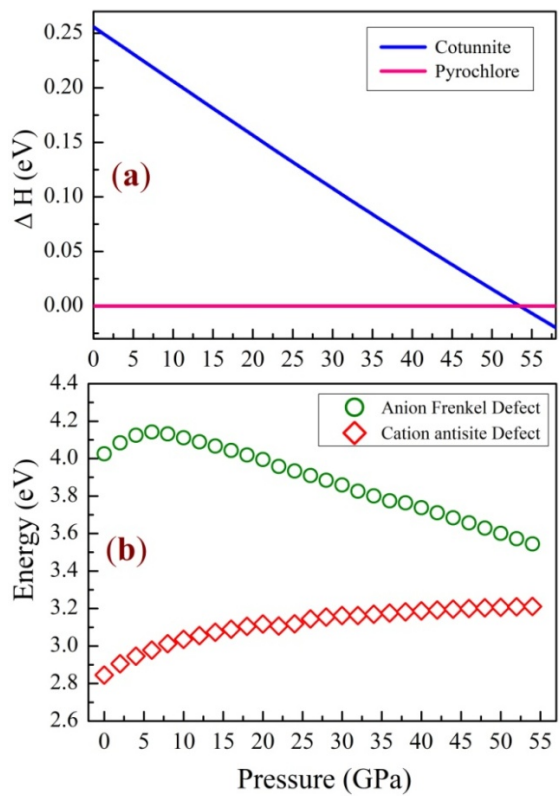

**Figure 10.** (a) Relative enthalpy of high-pressure cotunnite phase with respect to ambient pyrochlore phase. (b) Cation antisite defect and anion Frenkel defect formation energy in pyrochlore phase with respect to pressure.

Theoretical $P-V$ data has been fitted with the third order BM EoS and obtained values of B and B' for pyrochlore structure are 189 (12) GPa and 4 (3) respectively. Similarly, for the high-pressure cotunnite phase the corresponding values are 168 (10) GPa and 5 (2) respectively. It suggests that the high-pressure phase is more compressible than the pyrochlore phase.

The commonly accepted mechanism of structural transformation from pyrochlore to cotunnite phase[57] is in terms of the ease of point defects formation in the lattice mainly cation antisite defects and oxygen Frenkel defects[58-60].

Therefore, in order to study the mechanism of structural transformation from ordered pyrochlore to defect cotunnite phase we have also performed the defect formation energies *i.e.* cation antisite defects and anion Frenkel defect in the pyrochlore lattice with pressure. Cation antisite defect have been calculated by the difference of the total energies between optimized structure and relaxed defect structure with one $Er^{3+}$ ions exchanged with the nearest $Ti^{4+}$ ions[58]. The anion Frenkel defect formation energies have been calculated by the difference between the optimized structure and the relaxed defect structure moving one oxygen atom from *8b* to *8a* vacant site. The originally unoccupied *8a* site is considered as an interstitial site. Though, migration from *8b* to *8a* is expected to be unfavorable due to an intervening tetravalent cation and hence put an upper bound to the Frenkel defect formation energies. The other path for oxygen ion movement *i.e. 48f* to *8a* is expected to be less energy consuming[60]. The variation of defect formation energies with pressure have been shown in Fig. 10(b). It shows that cation antisite defect formation energy lies lower than the oxygen ion Frenkel defect. The system therefore prefers to first generate cation antisite defects. Thereafter, the formation of anion Frenkel defects becomes much more favorable with the presence of cation antisite defects[58]. In addition, the value of the anion Frenkel defect in our system is lower than many other pyrochlore compounds reported in the literature. Fig. 10(b) also indicates that the formation energy of Frenkel defects decreases with pressure while energy of cation antisite defects remains almost unchanged at higher pressure. Hence the pyrochlore lattice of $Er_2Ti_2O_7$ can accommodate the various defects and it can support the formation of disordered structure. It is important to mention here that if the lattice could not hold the point defects, then system would prefer to undergo amorphization. In high-pressure, the untransformed ambient phase consists large number of point defects, while releasing pressure it cannot able to accommodate them leading to their amorphization.

In general the defect formation energy in the pyrochlore lies in order of a few eV. When a system is compressed inside DAC, then the mechanical work done is given as $\sim \int PdV$, P is pressure and $dV$ is the change in volume. In our system, phase transition pressure $P \sim 50$ GPa, at this pressure, from the DFT calculations we know $\Delta V \sim 170$ Å$^3$. From the $P - V$ curve, the total work done per formula unit comes about $\sim 2.8\ eV$. Hence through the compression, sufficient energy is supplied to system, which is required for generation of point defects.

## CONCLUSION

We have observed that the ambient pyrochlore $Er_2Ti_2O_7$ undergoes irreversible structural phase transition initiating at ~40 GPa. The high-pressure orthorhombic cotunnite phase coexists with the ambient phase till the highest pressure of the experiment ~ 60.0 GPa. During the phase transition there is volume collapse of about ~11.2% experimentally indicating that the transformation is first order in nature. After complete release of pressure, the high-pressure cotunnite structure sustains, while the fraction of untransformed pyrochlore phase becomes amorphous. High-pressure Raman and photoluminescence studies also agree with XRD and suggest local site symmetry lowering under pressure after the structural transition. The local coordination geometry of the recovered sample has been probed using x-ray absorption experiments, and it is found that high-pressure phase has nine coordinated $Er^{3+}$ ions. DFT calculations also support experimental findings that cotunnite phase is the favorable high-pressure phase above 53 GPa and pyrochlore lattice can accommodate points defect, necessary in order to undergo order-disorder phase transition.

**Acknowledgement:** Authors would like to thank Dr. V. Srihari for user support in high-pressure x-ray diffraction at beamline-11, INDUS-2. Authors are also grateful to Dr. A. K. Mishra, Dr N. N. Patel for useful discussions and Dr. T. Sakuntala for continuous support and encouragements.